# T-A homogenized and multi-scale models with real-time simulation capabilities for large-scale HTS systems

Edgar Berrospe-Juarez[1], Víctor M R Zermeño, Frederic Trillaud[3] and Francesco Grilli[2]

[1]Postgraduate School of Engineering, National Autonomous University of Mexico, Mexico

[2]Karlsruhe Institute of Technology, Germany

[3]Institute of Engineering, National Autonomous University of Mexico, Mexico

email: eberrospej@iingen.unam.mx

This work was supported in part by the Programa de Maestría y Doctorado en Ingeniería of the Universidad National Autónoma de México (UNAM) and the Consejo Nacional de Ciencia y Tecnología (CONACYT) under CVU: 490544, and by DGAPA-UNAM grant, PAPIIT-2017 #TA100617.

**Abstract:** The emergence of second-generation high temperature superconducting tapes has favored the development of large-scale superconductor systems. The mathematical models for superconductors have evolved from simple analytical models to complex numerical models. The available analytical models are just capable of estimating electromagnetic quantities in single wires or small arrays. The numerical models like the *H* formulation finite element models are useful for the analysis of more medium-size systems, but their application in large-scale systems is problematic due to the excessive cost in terms of memory and the computation time. Then it is necessary to devise new strategies to tackle this challenge. The homogenization and the multi-scale approach are methodologies that have successfully simplified the description of the systems allowing the study of large-scale systems. Recently, efficient calculations have been achieved using the *T-A* formulation. In this manuscript, we analyze the adequate order of the elements used in the *T-A* formulation. More importantly, we propose a series of adaptations to the homogenization and multi-scale methodologies in order to allow their implementation to the *T-A* formulation models. The proposed strategies are validated by means of comparing them with the *H* reference model of a racetrack coil. The computation time and memory demand are substantially reduced up to a point that makes it possible to perform real-time simulations for slow ramping cycles. At the same time the comparison demonstrates a very good agreement with respect to the reference model, both at a local and global level, is achieved.

# 1. Introduction

The high temperature superconductors (HTS) materials have experienced an important progress during the recent years. Nowadays, the high current capacity of second-generation (2G) HTS tapes has made possible the production and commercialization of superconducting fault current limiters and cables [1]. The community is also interested in developing other systems based on 2G HTS tapes like generators, motors, power transformers [2], and also high field scientific magnets [3-5]. These devises are called large-scale superconductor systems, because they are made from hundreds to thousands of turns of HTS tapes. The design and operation of large-scale superconductor systems requires the availability of mathematical models capable of estimating the hysteresis losses and other electromagnetic and thermal quantities. In particular, fast numerical simulations can propel the development of devises allowing parametric analysis.

The first available analytical models were limited to the analysis of one single conductor [6-8]. More complex models considering conductor stacks under not general conditions are presented in [9-12]. The reader can refer to [13] for a deep a review of available analytical models. The numerical models open the possibility of considering systems with larger number of tapes. Methods like the finite element method (FEM) are well documented in the literature [14-16]. Nevertheless, the superconductor systems have specific characteristics, that should be considered in order to successfully address the modelling task [17]. For a deeper review of numerical models for HTS, the reader can refer to [18] and [19].

The FEM models of HTS tapes use different formulations of the Maxwell's equations. The differences between the formulations comes from the variety in the state variables. The first proposed FEM model of an HTS tape uses the $T$-$\varphi$ formulation [20], a later model uses the $A$-$V$ formulation [21]. Other formulations are compiled in [17] and [22]. During the last years the $H$ formulation [22] has become a de facto standard in the community. In the recent publications [23] and [24] the $T$-$A$ formulation has been described, this formulation couples together the $T$ and $A$ formulations. The FEM models using the $T$-$A$ formulation allows to address the analysis of large-scale systems. As a prelude for the strategies that are going to be proposed in this manuscript, we address the analysis of the order of the elements used in the $T$-$A$ formulation. The analysis is conducted by means of performing different simulations using different order of elements.

The main concern when dealing with large-scale systems is the large number of turns. The homogenization and the multi-scale approaches can considerably reduce the description of the systems without compromising accuracy of the results. The homogenization process consists in the assumption that a tapes stack can be described by an anisotropic bulk. This approach was first proposed in [25] and later successfully improvements have been reported in [26-28]. The multi-scale approach is based on the idea to simulate a reduced number of tapes, the interaction between tapes is achieved by means of the boundary conditions containing the background magnetic field produced by the rest of the tapes. This method was proposed in [29], with later refinements in [30] and an iterative version proposed in [31]. Among the available methodologies suitable for the analysis of large-scale systems we should mention the Minimum Magnetic Energy Variation (MMEV) method proposed in [32] and [33]. The MMEV method evolved into the Minimum Electro-Magnetic Entropy Production (MEMEP) method [34], [35].

In this manuscript, we present two strategies to address the numerical modelling large-scale superconductor systems. These new strategies arose from the study of the implementation of homogenization and multi-scale methodologies together with the *T-A* formulation models. The application is not straightforward and some delicate adaptations are required. It is also demonstrated that the *T-A* homogenized and *T-A* multi-scale models are so efficient that it is possible to achieve real-time simulations of the case study using a desktop computer.

This manuscript is organized as follow. Section 2 starts with the description of the *T-A* formulation, then the *T-A* homogenized and *T-A* multi-scale strategies are presented. Section 3 contains the description of the case study racetrack coil as well as its *T-A* full model. This section ends with the study of the adequate choice for the order of the elements used in the *T-A* models. The description of new proposed *T-A* homogenized and multi-scale models of the case study is presented in section 4. Sections 5 contains the simulation results and the comparison of the different models. An important conclusion about the possibility of real-time simulations is reached in section 6. Finally, the conclusions are exposed in section 7.

## 2. Modelling Methodology

**T-A Formulation**

The *T-A* formulation was reported in [20] and [21]. The *T-A* formulation approach solves two state variables, the current vector potential $\boldsymbol{T}$ along the HTS layer, and the magnetic vector potential $\boldsymbol{A}$ all over the whole geometry. The *T-A* formulation assumes that the HTS layer of the tape have an infinitesimal thickness. In a 2D geometry this assumption means that the HTS layers are assumed to be 1D lines. Then, the $\boldsymbol{T}$ and $\boldsymbol{A}$ vectors have only one component.

The magnetic vector potential $\boldsymbol{A}$ is defined by

$$\boldsymbol{B} = \nabla \times \boldsymbol{A} \tag{1}$$

The governing equations of the *A* formulation is

$$\nabla \times \nabla \times \boldsymbol{A} = \mu \boldsymbol{J} \tag{2}$$

where $\mu$ is the magnetic permeability.

The current vector potential $\boldsymbol{T}$ is defined by

$$\boldsymbol{J} = \nabla \times \boldsymbol{T} \tag{3}$$

The governing equation of the *T* formulation is

$$\nabla \times \rho \nabla \times \boldsymbol{T} = -\frac{\partial \boldsymbol{B}}{\partial t} \qquad (4)$$

where $\rho$ is the resistivity.

In the 2D case depicted in figure 1, equations 3 and 4 are reduced to

$$J_z = \frac{\partial T_y}{\partial x} \qquad (5)$$

$$\frac{\partial}{\partial x} \rho \frac{\partial T_y}{\partial x} = \frac{\partial B_y}{\partial t} \qquad (6)$$

The necessary boundary conditions can be found considering the transport current in the tape

$$I = \iint_S \boldsymbol{J}\, dS = \iint_S \nabla \times \boldsymbol{T}\, dS = \oint \boldsymbol{T}\, ds \qquad (7)$$

$$I = (T_1 - T_2)\delta \qquad (8)$$

where $\delta$ is the thickness of the HTS layer. $T_1$ and $T_2$ are shown in figure 1.

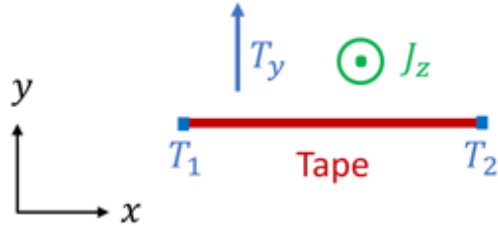

**Figure 1.** Infinitesimal thickness HTS tape in a 2D geometry.

The time derivative of the magnetic field $\boldsymbol{B}$, obtained by calculating $\boldsymbol{A}$, is used by the $T$ formulation. At the same time, the current density $\boldsymbol{J}$, obtained by calculating $\boldsymbol{T}$, is multiplied by the thickness of the HTS, and then forced into the $A$ formulation as an external surface current density.

The models that consider each tape inside the system, and applies the *T-A* formulation as proposed in [21] and briefly described here, are called *T-A full models* in the rest of this manuscript.

## Homogenization

The homogenized models assume that an HTS tapes stack can be modeled as an anisotropic bulk, which is a simpler object to simulate. This simplification in the geometric description of the stack should not compromise its electromagnetic behavior, thus the model construction requires additional considerations.

As well as in the *T-A* full models, when the *T-A* formulation is applied to a homogenized model, the influence of the component of **B** parallel to the surface of the tape is neglected, and **T** has only one component, the same as in the *T-A* formulation models, see equation (6). Figure 2 illustrates the homogenization process. The bulk brings together all the unit cells containing the HTS tapes. The governing equation of the bulk domain is also (6), and the only state variable is $T_y$. The same boundary conditions from equation (8) are applied to the edges instead to the extreme points. The bulk with these boundary conditions can be understood as the limiting case of a densely packed stack made up of tapes with infinitesimal thickness. The resistivity inside the bulk is considered to be the resistivity of the HTS material, thus the strategy proposed here does not consider the copper and other normal conductors that constitute the HTS tapes. The homogenization process is depicted in figure 2, together with the boundary conditions applied in the homogenized bulk.

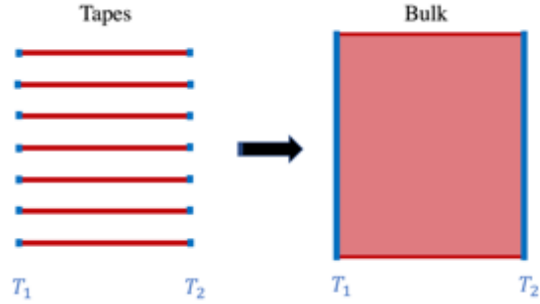

**Figure 2.** Homogenization process and boundary conditions for the *T-A* homogenized bulk. The stack is transformed in a bulk. The boundary conditions are applied to the edges corresponding with the extreme points of the tapes.

Conversely to the non-homogeneous *T-A* model, here the **J**, obtained by calculating **T** inside the bulk, is multiplied by the ratio $\delta/\Lambda$, and forced into the *A* formulation as an external current density. Where $\Lambda$ is the thickness of the unit cell. It is important to emphasize that the *A* formulation is applied all over the full geometry including the air domains, while the *T* formulation is applied just in the domains inside the homogenized bulks.

**Multi-scale approach**

The multi-scale and the iterative multi-scale method as described in [30] and [31], respectively use two different submodels. The single tape submodel is a *H* formulation model that computes the *J* distribution in the analyzed tapes. The coil submodel is an *A* formulation model that computes the background magnetic field. The two submodels run separately, then the *J* distribution and the background field are not computed simultaneously.

The multi-scale approach proposed here is also based in the *T-A* formulation. As in the *T-A* full models, there is only one model and the *T* and *A* formulations are coupled together. Hence, these new multi-scale models simultaneously compute the *J* distribution in the analyzed tapes and the background field produced by all the tapes. The multi-scale *T-A* models consider a reduced number of analyzed tapes, each analyzed tape is modeled in detail using the *T-A* formulation model. The remaining non-analyzed tapes and the surrounding air are modeled using just the *A* formulation. The *J* distributions in the non-analyzed tapes are approximated by linear interpolation of the *J* distributions in the analyzed tapes. The interpolated *J* distributions in the non-analyzed tapes are also multiplied by the thickness of the HTS layer, and forced into the *A* formulation as surface current densities. Figure 3 shows the multi-scale approach applied in a small stack.

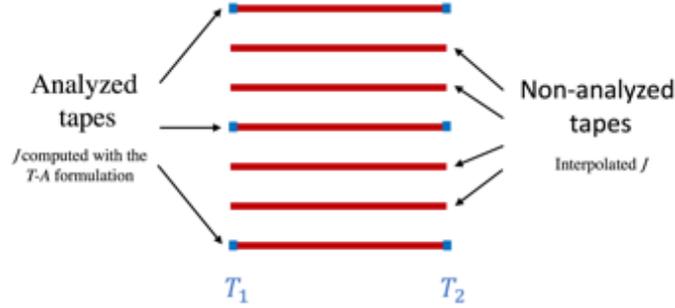

**Figure 3.** Multi-scale approach. In this example, there are 3 analyzed tapes and the *J* distribution in the non-analyzed tapes is approximated by linear interpolation.

## 3. Case study, *H* reference and *T-A* models

**Case study**

The case study used in this manuscript is the same racetrack coil used in [30]. The coil has 10 pancakes (stacks), each composed of 200 turns. The symmetry allows to model just one quarter of the system, this means it is possible to consider 5 pancakes each one with 100 tapes.

The *H reference model* of this manuscript is also the same reference model used in [30]. This model uses the *H* formulation and consider each tape. The regions containing just one HTS layer and its surrounding air are called unit cells. The mesh of the unit cells is structured, and considers 1 element along the tape's thickness and 100 elements along the tape's width. The 100 elements are distributed symmetrically with an increasing number of elements at the extremities of the tape. The HTS layers are considered to be surrounded with air, thus the copper and other normal conductors that constitute the HTS tapes are not considered.

The electrical resistivity of the HTS material is modeled by the so-called *E-J* power-law [36],

$$\rho_{HTS} = \frac{E_c}{J_c(\boldsymbol{B})} \left| \frac{\boldsymbol{J}}{J_c(\boldsymbol{B})} \right|^{n-1}. \tag{9}$$

The critical current density $J_c$ is defined by a modified Kim's relation [37], this relation describes the anisotropic dependence of the on the magnetic field

$$J_C(\boldsymbol{B}) = \frac{J_{c0}}{\left(1 + \frac{\sqrt{k^2 B_\parallel^2 + B_\perp^2}}{B_0}\right)^\alpha},\tag{10}$$

where $B_\perp$ and $B_\parallel$ are the magnetic field components perpendicular and parallel to the wide surface of the tape. Since there are no magnetic materials are considered, the permeability of the air and the HTS material is chosen to be the permeability of the vacuum $\mu_0$. The parameters of racetrack coil are resumed in table 1.

**Table 1.** Racetrack coil parameters.

| Parameter | Value |
|---|---|
| Pancakes | 10 |
| Turns per pancake | 200 |
| Unit cell width | 4.45 mm |
| Unit cell thickness | 293 μm |
| HTS layer width | 4 mm |
| HTS layer thickness | 1 μm |
| $E_c$ | 1e-4 Vm$^{-1}$ |
| $n$ | 38 |
| $J_{c0}$ | 2.8e10 Am$^{-2}$ |
| $B_0$ | 0.04265 T |
| $k$ | 0.29515 |
| $\alpha$ | 0.7 |
| Air resistivity | 1 Ωm |

### *T-A* Full Models

The *T-A* full models consider in detail all the tapes. The mesh used in the tape's region is a structured mesh. To allow an assessment of the number of degrees of freedom (DOF) and the accuracy of the results, three models with different numbers of elements along the tape's width were built. The first model uses 100 elements distributed as in the reference model. The other two models use 50 and 25 elements along the tape's width, respectively, both with a uniform distribution of elements. It is also important to note that the *T-A* full models use Lagrange elements.

The geometry and the mesh of the model with 100 elements along the tape's width is shown in figure 4. This figure also shows the numbering of the tapes and pancakes.

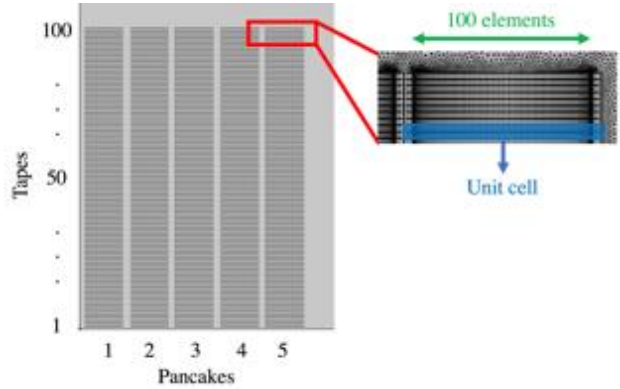

**Figure 4.** Geometry and mesh of the *T-A* full model. The analyzed section of the case study has 5 pancakes with 100 tapes per pancake. The mesh in the unit cells is structured.

## Spurious oscillations in the *J* distributions

A comprehensive presentation of the simulations results of the *T-A* full models is going to be presented in section 5, in conjunction with the results of the homogenized and multiscale models. Here, we present in advance of the *J* distributions obtained with the *T-A* full model. These results will justify the election of the orders of the elements in the rest of the models.

The references model and the *T-A* full model with 50 elements along the tape's width were simulated for one cycle of a 11 A, 50 Hz transport current. The *T-A* model was simulated three times using different order elements. The first simulation uses first order elements for both *T* and *A* formulations, the second simulation uses second order elements for both formulations, the last simulation uses first order elements for the *T* formulation and second order elements for the *A* formulation. The *J* distribution of the tape 25 of pancake 5, at $t=2$ ms are shown in figure 5.

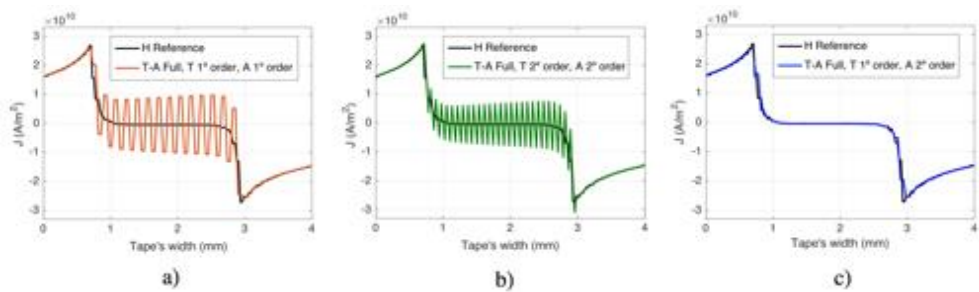

**Figure 5.** *J* distribution of the tape 25 of pancake 5 at $t=2$ ms. The simulations were performed with the *H* formulation reference model and with the *T-A* full model with 50 elements along the tape's width. a) The *T-A* full model uses first order elements for both formulations. b) The *T-A* full model uses second order elements for both formulations. c) The *T-A* full model uses first order elements for the *T* formulation and second order elements for the *A* formulation.

The *J* distributions of the *T-A* models look similar to the *J* distributions of the reference model. But, the results show that the *J* distributions of the *T-A* models present spurious oscillations. Even though, the reference [24] contains a discussion about the increment in the computation time due to the increment of the order of the elements, these oscillations are not discussed. The spurious oscillations are present for subcritical current densities. It is possible to observe that the "frequency" of the oscillations when both formulations uses second order elements is twice the frequency when both formulations use first order elements. The *T-A* full model with 50 elements was chosen for this assessment because it is easier to appreciate the oscillations.

The oscillations disappear when the order of the elements is higher in the *A* formulation than in the *T* formulation. Hence, throughout the rest of this manuscript it is assumed that the *T-A* full models are models using first order Lagrange elements in the *T* formulation and second order Lagrange elements in the *A* formulation, unless otherwise indicated. Similarly, the homogenized and multiscale models, described in the next section, use this combination of order elements to avoid the oscillations.

The problem of spurious oscillations is known to appear in other fields of physics like the fluid mechanics. There are finite element spaces that are unstable for the Navier-Stokes equations. A stable pair of elements should be used when both the velocity and the pressure are required, otherwise unwanted oscillations may be present in the pressure solution. One simple solution is the use of the Taylor-Hood elements which use first order elements for the pressure and second order elements for the velocity, see [38].

## 4. T-A homogenized and multi-scale models

**T-A homogenized models**

The *T-A* homogenized model of the case study consider 5 bulks, each per pancake, and a structured mesh with 6 elements along the bulk's width. In order to permit a fair comparation with the *T-A* full models, three models were built. The models use 100, 50 and 25 elements along the tape's width, with the same symmetric and uniform distributions used in the *T-A* full models. The geometry and the mesh of the model with 100 elements along the tape's width is shown in figure 6. The mesh has more elements in the upper part of the pancakes where higher losses and screening currents are expected. Contrarily, the central parts of the pancakes require less elements.

Considering the undesired ripples produced when first order elements are used for both formulations, the homogenized models use first order elements for the *T* formulation and second order elements for the *A* formulation.

The hysteresis losses are computed in the center of each of the 6 elements along the bulks. These losses, corresponding to the tapes located at the center of the 6 elements, are used to estimate the losses in the rest of the tapes. The interpolation method used in this manuscript is the Piecewise Cubic Hermite Interpolating Polynomial (PCHIP) method [39].

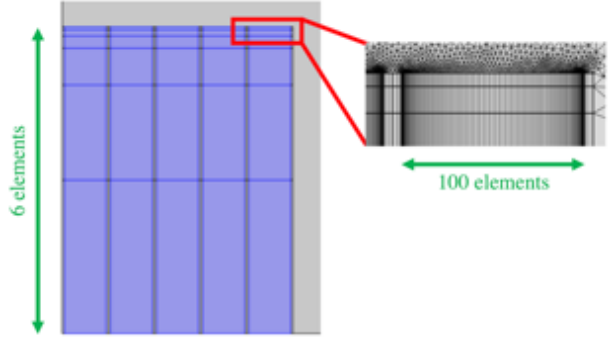

**Figure 6.** Geometry and mesh of the *T-A* homogenized model. The bulk's mesh considers 6 elements along the bulks, with a denser distribution in the upper part of the pancakes

*T-A* **multi-scale models**

The multi-scale models in this manuscript consider 30 analyzed tapes, the election of the analyzed tapes follows the same directives proposed in [31]. The set of analyzed tapes in each pancake is {25, 66, 88, 96, 99, 100}. This distribution allows to reproduce the expected variations in the losses at the extreme parts of the pancakes.

For the purpose of avoiding the undesired oscillations, the unit cells of the analyzed tapes use first order elements for the *T* formulation and second order elements for the *A* formulation. The unit cells of tapes closest non-analyzed tapes also use second order elements for the *A* formulation, while the rest of the system first order elements are used for the *A* formulation. The domains using first order elements in the *A* formulation are connected with the domains using second order elements in the *A* formulation by means of time dependent Dirichlet boundary conditions.

A structured mesh is used for the domains with the pancakes, and a triangular mesh is used for the surrounding air. The unit cells where the second order elements are used consider a mesh with more elements along the tape's width than the regions where first order elements. The transition zones between unit cells with different number of elements use triangular meshes. The position of the analyzed tapes, the regions with second order elements and the mesh are shown in figure 7, this figure correspond to a model with 100 elements along the tape's width in the regions with second order elements and 50 elements in the rest of the tapes. The second model uses a mesh with 50 and 25 elements for the tapes with second and first order elements, respectively. The last model uses a mesh with 25 and 13 elements. The distribution of the elements in the last two models is uniform, while the distribution of the elements in the firs model is the same symmetric distribution used in the homogeneous and full models with 100 elements along the tape's width.

In the multi-scale model, the PCHIP interpolation method is also used, to approximate the hysteresis losses in the non-analyzed tapes based in the losses of the analyzed tapes.

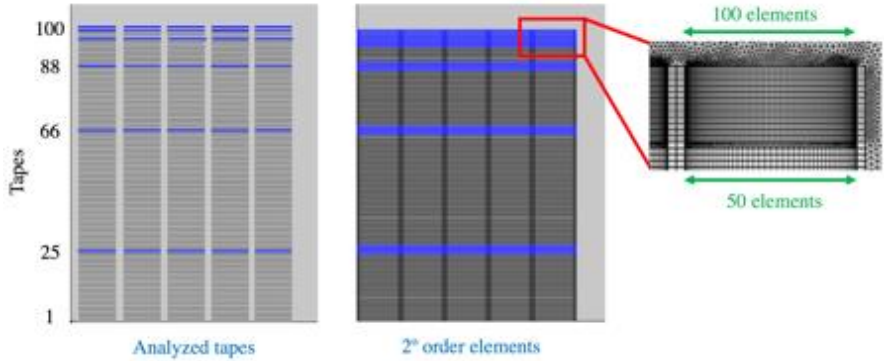

**Figure 7.** Analyzed tapes, regions with second order elements and the mesh of the *T-A* multi-scale model. The increased number of analyzed tapes in the upper part of the pancakes allows an accurate estimation of losses and screening currents

## 5. Results

The *H* reference model as well as the *T-A* full, T-A homogenized and *T-A* multi-scale were simulated for a 11 A, 50 Hz sinusoidal transport current. The results of the reference model are considered as the best estimation for the different electromagnetic quantities. Figure 8 show the results models with 100 elements along the tape's width. This figure allows to make a qualitative comparison between the models. The normalized current density ( $J_n = J/J_c$ ) and the magnetic flux density magnitude ( $|B|$ ), both at the negative peak current (*t*=15 ms), are shown in the first and second row. The *T-A* models can successfully reproduce the *J* distribution produced by the *H* reference model. The third $J_n$ plot show the $J_n$ all across the bulks, we decided to present the plot in this way to stress that it corresponds to the *T-A* homogenized model, but if we had decided to present a plot with the $J_n$ only across the tape domains, the plot would be also indistinguishable. As a direct consequence of the similarity between the *J* distributions produced by the four models, the magnetic field plots in figure 8 are identical at simple sight.

The third row in figure 8 show the average hysteresis losses in the coil. The x-axis in the plots represents the tapes inside every pancake. The average losses are calculated as the average of the losses in the tapes during the second half of the 50 Hz cycle. The losses in each pancake are plotted in different lines, one for each pancake. The losses in pancake 5 are almost three orders of magnitude larger than the losses in pancake 1. This difference is linked to the higher current penetration in the tapes across the pancakes, and to the difference in the field direction. The plots in the third row of figure 8 demonstrate that the three *T-A* models can achieve accurate estimations of hysteresis losses at local level.

The last row in figure 8 are the *J* distributions in tape 25 of pancake 5, at *t*=2ms. Thanks to the appropriate choice of the order of the elements, the J distribution estimated with the T-A models show no spurious oscillations. The *J* distribution of the *T-A* full model is more accurate than the one in figure 5 c), because it corresponds to the model with 100 elements along the tape's width, and the one in figure 5 correspond to the *T-A* full model with 50 elements. The *J* distribution of the *T-A* homogenized model shows two small peaks attached to the current fronts. These peaks are caused by the locally different **B** distribution across the bulk, this feature is also present in the *H* formulation homogenized models. Finally, the *J* distribution of the *T-A* multi-scale model is, as accurate as the one of the *T-A* full model. It is important to emphasize that the tape 25 of pancake 5 is one of the analyzed tapes in the *T-A* multi-scale model.

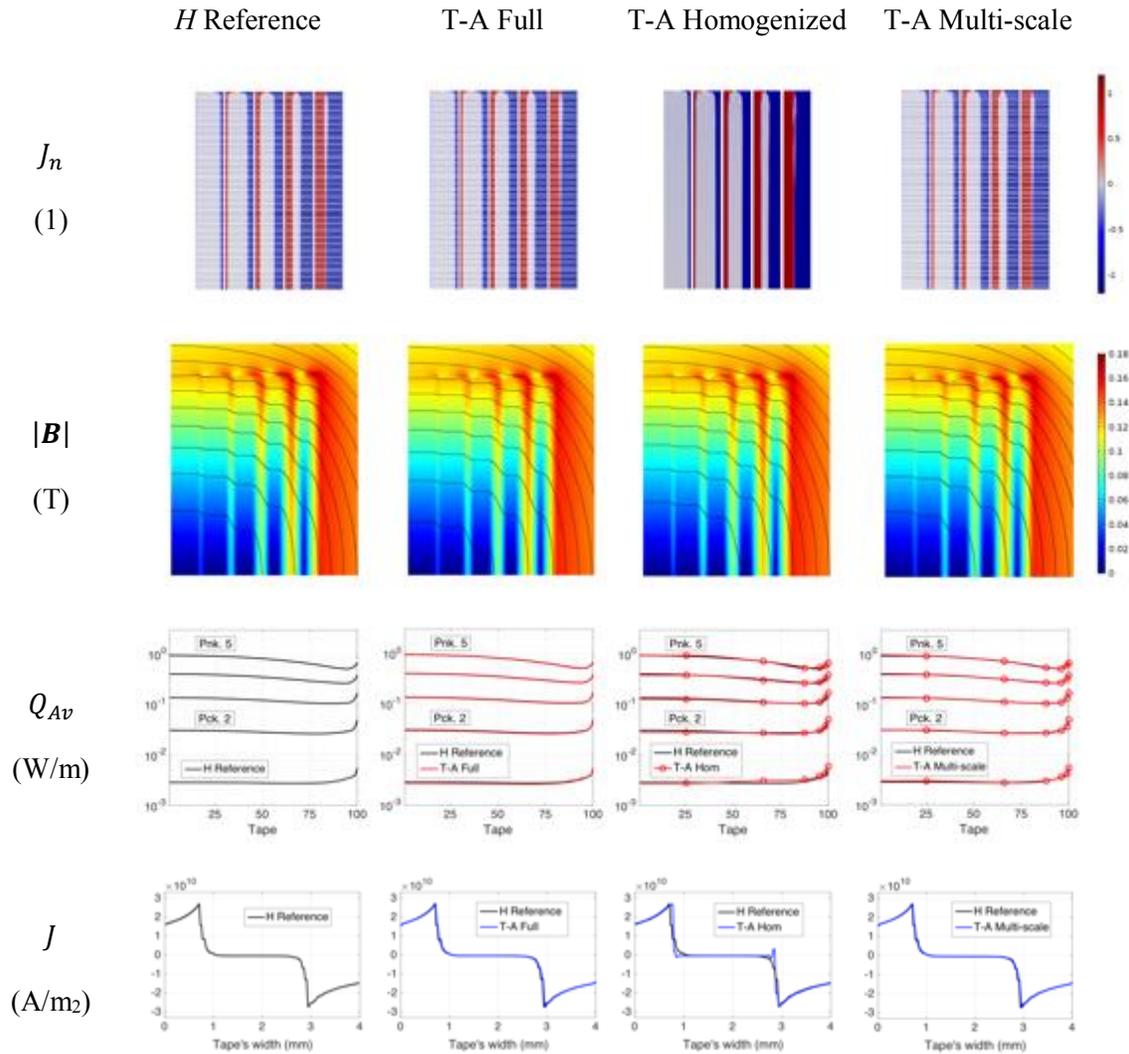

**Figure 8.** *H* reference and T-A models results with a 11 A, 50 Hz transport current. The first column shows the results of the reference model. The second, third and fourth row shows the results of *T-A* full, homogenized and multi-scale models, respectively. The plots for $J_n$ and $|\boldsymbol{B}|$ show the results at peak transport current, t =15 ms. The *J* plots show the results of the tape 25 of pancake 5, at $t=2$ ms.

## Quantitative comparison and number of elements assessment

Figure 9 present a quantitative comparison between the *T-A* models and the *H* reference model. The results are presented as function of the number of elements along the tape's width. The figure 9 a) shows the relative error in the total losses incurred by the *T-A* models when compared with the *H* reference model. Here the adjective total means that the losses in each tape are summed. It is remarkable that for 50 or more elements all the errors are less than 2 %. The total average losses of the *H* reference model are 127.24 W/m.

The coefficient of determination ( $R^2$ ) is a widely used metric to evaluate the goodness of fit [40], here is used to evaluate the accuracy of the *J* distributions.

$$R^2 = 1 - \frac{\sum_{i=1}^{m}(J_H - J_{TA})^2}{\sum_{i=1}^{m}(J_H - \overline{J_H})^2} \, , \tag{11}$$

where $J_{H/TA}$ are vectors containing the uniformly sampled *J* distribution of all the tapes, at all time steps (including those of the first half of the cycle), computed with the *H* or *T-A* models. The figure 9 b) shows the $R^2$ of the different models. The *T-A* multi-scale models are more suitable for the estimation of *J* distributions. The lower achieved $R^2$ value is the one of the *T-A* full model with first order elements for both *T* and *A* formulations. The origin of this low value is the presence of the spurious oscillations. Due to the same drawback, the $R^2$ of the *T-A* full model with second order elements for both formulations is lower than the $R^2$ of the *T-A* full (*T*-1º, *A*-2º). The spurious oscillations have no impact in the losses, because the oscillations occur for subcritical values of *J*, but correct orders should be chosen to achieve the best possible *J* distribution.

The DOF of the *H* reference model is 563893. Figure 9 c) shows the number of DOF of the T-A models, expressed as a percentage of the DOF of the *H* reference model. The DOF of the T-A homogenized models are approximately 20 times lower than those of the *T-A* full models, this is a clear picture of the reduction of the computational load achieved with the homogenization. Figure 9 d) shows the computation times expressed as a percentage of the computation time of the *H* reference model, which is 31 h 33 min. In the worst-case scenario, the *T-A* homogenized and multi-scale models require less than 10 % of the time required by the *H* reference model. The *T-A* full models are efficient [24], but the computation time of the *T-A* homogenized models are one order of magnitude lower. The computer used to perform the simulations is a MacBook (3 GHz Intel Core i7-4578U, 4 cores, 16 GB of RAM).

The DOF in the homogenized model can be further reduced by reducing the number of elements along the bulk. While, the DOF in the multi-scale model can be reduced, by reducing the number of analyzed tapes. It has been demonstrated in [31] that it is possible to reduce the number of analyzed tapes in the central pancakes without losing accuracy in the total losses,

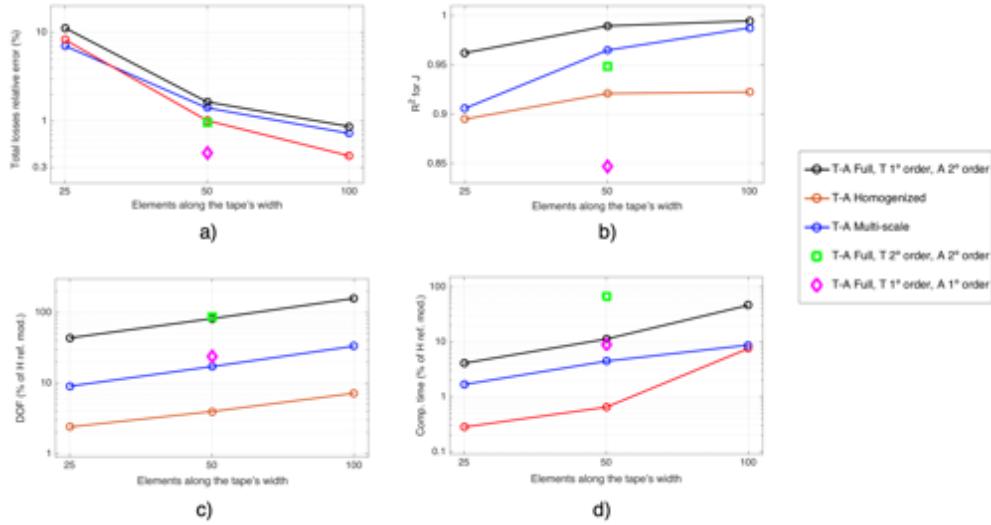

**Figure 9.** Comparison between the *T-A* models and the *H* reference model. The T-A models are tested for different number of elements along the tape's width. a) Absolut relative total losses error. b) Coefficient of determination of *J*. c) DOF expressed as a percentage of the DOF of the *H* reference model. d) Computation time expressed as a percentage o the computation time of the *H* reference model.

**Time required to build the models**

All the models described in this manuscript, were implemented in COMSOL Multiphysics 5.3. One important parameter is the time required to build the models. This is a quantity difficult to measure because it strongly depends on the expertise of the analyzer. The geometry and the mesh in the *H* reference model are not so easy to build, it is necessary to build the unit cells and then generate an array. The definitions of the integral constraints that impose the transport current in each tape is a time-consuming process, fortunately this process can be programed using Matlab, then the time required to build the *H* reference model is around 1 h. The process to build the *T-A* full model is easier, once the unit cell is ready, it can be copied to produce and array that already includes the boundary conditions. The time required to build the *T-A* full model is less than 30 min. Similarly, the *T-A* homogenized model is easy to build, instead of a unit cell the analyzer needs to create a bulk and then copy the bulk. A smaller number of boundary conditions is required for the *T-A* homogenized model, anyway we consider that the *T-A* homogenized model requires less than 30 min for its construction. Finally, the process to build the *T-A* multi-scale model requires around 3 h. This large time is required to build the unit cell, create and array of unit cell, select the analyzed tapes, set the properties of the analyzed tapes as well as the properties of the non-analyzed tapes, and define the functions required to implement the interpolation of the current densities.

## 6. Real-time capabilities

The *H* reference model does not allow to simulate the coil under slow transport currents. For example, the charging/discharging cycles, with periods in the order of the hours, that are present in real operation of magnets [4]. When the frequency of the transport current in the *H* reference model is changed from 50 Hz to a 5e-4 Hz, the time steps of the simulation cannot be increased at the same rate, and the computation time required to simulate one single cycle become prohibitive. In contrast, the models based on the *T-A* formulation does not have this restriction.

The *T-A* models with 100 elements along the tape's width were simulated for the ramping cycle show in blue line in figure 10, 1 h ramping up, plus 30 min plateau, plus 1 h ramping down, plus 30 min plateau. The losses as a function of time are also shown in the figure 10. The total losses during the 10800 s of the analyzed lapse and the computation time are summarized in table 2. As far as the *H* reference model was not simulated under these conditions, the losses estimated with the *T-A* full model are considered to be the best approximation. The losses estimated the *T-A* multi-scale model are more accurate than those estimated with the homogenized model, nevertheless the error with the homogenized model is low with a value of 1.6 %. The most important result is the computation time achieved with the *T-A* homogenized and multi-scale models, both times are less than the actual physical time of the simulation. Thus, the *T-A* homogenized and multi-scale models are suitable for real-time similations, feature that is not shared with the *T-A* full model, nor with the *H* reference model.

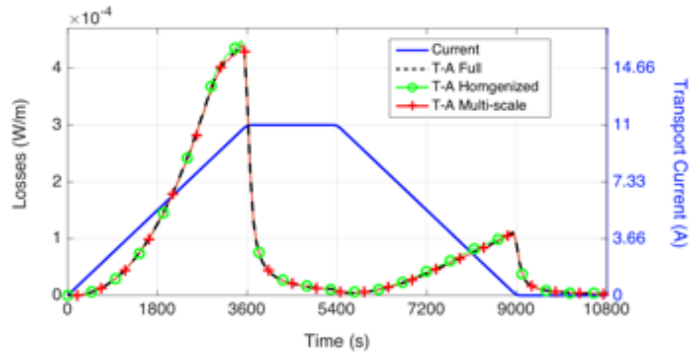

**Figure 10.** Infinitesimal thickness tape in a 2D geometry.

**Table 2.** Ramping cycle simulations.

| Model | Losses (J/m) | Computation Time |
|---|---|---|
| T-A full | 0.8772 | 39h 15min 43s |
| T-A Homogenized | 0.8916 | 1h 8min 20s |
| T-A Multi-scale | 0.8708 | 2h 35min 18s |

## 7. Conclusions

The *T-A* formulation reduces the geometric description of a 2D HTS tape with a large aspect ratio to a 1D line. The two new strategies proposed in this manuscript, the *T-A* homogenized and the *T-A* multi-scale models, are based in the *T-A* formulation, and also successfully take benefit of the homogenization and multi-scale methods. The homogenization was previously used with *H* formulation models, here we have investigated the necessary considerations to apply the homogenization together with the *T-A* formulation. This new strategy allows to build faster homogenized models.

The main limitation of the multi-scale method described in [30], is the absence of a clear definition of the *J* distribution. The iterative multi-scale method [31] overcomes this limitation. The *simultaneous* multi-scale method described in this manuscript can be seen as enhanced version of the iterative multi-scale method. The simultaneous multi-scale models are easier to construct than the iterative ones, because there is only one model instead of two submodels. The *T-A* multi-scale models can run faster because just one dynamic simulation is required, instead of the iterative implementation of *n* simulations.

The analyzes conducted in this manuscript show that the assumptions considered in the *T-A* homogenized and multi-scale models significantly reduce the DOF. These simplifications in the description of the systems does not compromise the accuracy of the results, and not only it is possible to accurately estimate the losses, but also the *J* distribution. The availability of the *T-A* homogenized and multi-scale models opens the possibility for the real-time simulation of large-scale superconductor systems. At the best of our knowledge, this last quality is new in the analysis of such complex systems.